\documentclass[sigconf]{acmart}

\AtBeginDocument{%
  \providecommand\BibTeX{{%
    \normalfont B\kern-0.5em{\scshape i\kern-0.25em b}\kern-0.8em\TeX}}}

\copyrightyear{2023}
\acmYear{2023}
\setcopyright{acmlicensed}\acmConference[KDD '23]{Proceedings of the 29th
ACM SIGKDD Conference on Knowledge Discovery and Data Mining}{August
6--10, 2023}{Long Beach, CA, USA}
\acmBooktitle{Proceedings of the 29th ACM SIGKDD Conference on Knowledge
Discovery and Data Mining (KDD '23), August 6--10, 2023, Long Beach, CA,
USA}
\acmPrice{15.00}
\acmDOI{10.1145/3580305.3599881}
\acmISBN{979-8-4007-0103-0/23/08}

\DeclareMathOperator*{\argmin}{argmin}
\usepackage{subcaption}
\usepackage{amsmath}

\begin{document}

\title{Optimizing Airbnb Search Journey with Multi-task Learning}

\author{Chun How Tan}
\orcid{0000-0002-2316-1259}
\affiliation{%
  \institution{Airbnb Inc.}
  \country{USA}
}
\email{chunhow.tan@airbnb.com}

\author{Austin Chan}
\orcid{0009-0002-8616-6986}
\affiliation{%
  \institution{Airbnb Inc.}
  \country{USA}
}
\email{austin.chan@airbnb.com}

\author{Malay Haldar}
\orcid{0009-0005-5128-0254}
\affiliation{%
  \institution{Airbnb Inc.}
  \country{USA}
}
\email{malay.haldar@airbnb.com}

\author{Jie Tang}
\orcid{0009-0006-9509-2367}
\affiliation{%
  \institution{Airbnb Inc.}
  \country{USA}
}
\email{jie.tang@airbnb.com}

\author{Xin Liu}
\orcid{0009-0008-7563-0425}
\affiliation{%
  \institution{Airbnb Inc.}
  \country{USA}
}
\email{xin.liu@airbnb.com}

\author{Mustafa Abdool}
\orcid{0009-0006-5454-075X}
\affiliation{%
  \institution{Airbnb Inc.}
  \country{USA}
}
\email{moose.abdool@airbnb.com}

\author{Huiji Gao}
\orcid{0009-0006-0424-248X}
\affiliation{%
  \institution{Airbnb Inc.}
  \country{USA}
}
\email{huiji.gao@airbnb.com}

\author{Liwei He}
\orcid{0009-0007-2942-3145}
\affiliation{%
  \institution{Airbnb Inc.}
  \country{USA}
}
\email{liwei.he@airbnb.com}

\author{Sanjeev Katariya}
\orcid{0009-0008-1519-174X}
\affiliation{%
  \institution{Airbnb Inc.}
  \country{USA}
}
\email{sanjeev.katariya@airbnb.com}

\renewcommand{\shortauthors}{Chun How Tan, et al.}

\begin{abstract}
  At Airbnb, an online marketplace for stays and experiences, guests often spend weeks exploring and comparing multiple items before making a final reservation request. Each reservation request may then potentially be rejected or cancelled by the host prior to check-in. The long and exploratory nature of the search journey, as well as the need to balance both guest and host preferences, present unique challenges for Airbnb search ranking. In this paper, we present Journey Ranker, a new multi-task deep learning model architecture that addresses these challenges. Journey Ranker leverages intermediate guest actions as milestones, both positive and negative, to better progress the guest towards a successful booking. It also uses contextual information such as guest state and search query to balance guest and host preferences. Its modular and extensible design, consisting of four modules with clear separation of concerns, allows for easy application to use cases beyond the Airbnb search ranking context. We conducted offline and online testing of the Journey Ranker and successfully deployed it in production to four different Airbnb products with significant business metrics improvements.
\end{abstract}

\ccsdesc[500]{Computing methodologies~Multi-task learning}
\ccsdesc[500]{Computing methodologies~Neural networks}
\ccsdesc[500]{Information systems~Learning to rank}
\ccsdesc[500]{Information systems~Content ranking}
\ccsdesc[300]{Information systems~Personalization}
\ccsdesc[500]{Applied computing~Online shopping}

\keywords{Search Ranking, Recommender Systems, User Search Journey, Multi-task learning, Two-sided marketplace}


\maketitle

\section{Introduction}
\label{section:introduction}
At Airbnb, search is the main way for a guest to discover the right inventory, such as stays, (in-real-life) experiences (i.e. activities to do), online (i.e. virtual) experiences, etc. Specifically, given query parameters from the guest (e.g. location, date range, etc), the Airbnb Search Ranking model ranks all eligible inventories to optimize for the desired business metric. While our proposed model architecture in this paper is applicable to other domains (e.g. four different product use cases within Airbnb), we will focus the majority of our discussion in this paper on one particular instantiation of the model architecture, for the Airbnb's stays product use case.

Unlike other products such as social media feeds, which focus on engagement, Airbnb focuses more on the final guest conversion, similar to other e-commerce and two-sided marketplaces. Furthermore, since booking a place to stay is generally a large expense, guests tend to consider a lot of options carefully, akin to purchasing a big-ticket item such as an electronic appliance, car, or house. Based on internal data, a guest could spend multiple weeks, across dozens of searches, before narrowing down options and making a final reservation request on Airbnb. Lastly, as a two-sided marketplace, the guest's reservation request may end up being rejected or cancelled by the host, resulting in a poor guest experience where the guest must start the whole search journey again. 

\begin{figure}
  \centering
  \includegraphics[width=\linewidth]{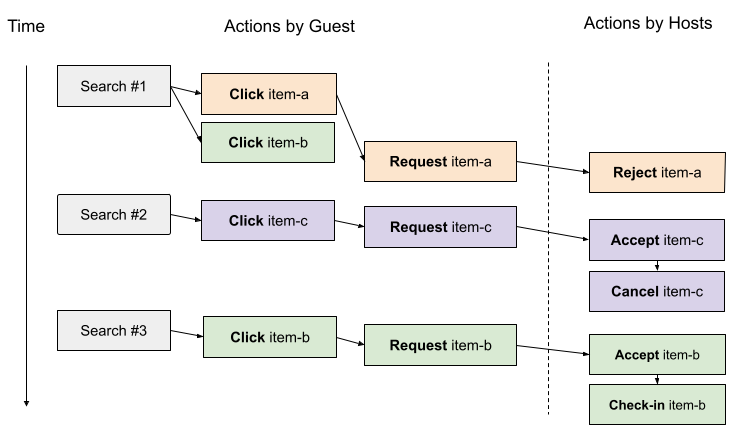}
  \Description{An example Airbnb guest journey who makes multiple searches, clicks, and reservation requests, before finally resulting in an uncancelled booking and check-in.}
  \caption{An example Airbnb search journey for a guest who makes multiple searches, clicks, and reservation requests, before finally resulting in a realized stay.}
  \label{figure:airbnb_journey_flowchart}
  \vspace{-4mm}
\end{figure}

Figure \ref{figure:airbnb_journey_flowchart} shows an example guest search journey from a search to a realized reservation check-in. Note that the guest made reservation requests for two listings but ended up being cancelled or rejected by the host, highlighting the importance of considering host preferences. Furthermore, even though the guest clicked on \textit{item-b} early in the search journey, the guest only made a reservation request for \textit{item-b} after more searches, illustrating the long nature of search journeys in the Airbnb product use case.

Thus, Airbnb Stays Ranking needs to guide the guest throughout the long search journey while also balancing both guest and host preferences so that all the stakeholders are satisfied with the final Airbnb stay. However, the exploratory and assorted guest activities at Airbnb present unique challenges in modeling the end-to-end guest search journey. Guests go through different stages, making the modeling problem no longer a single task with binary positive / negative signals. Instead, we need to capture sparse guests and hosts preferences at multiple stages (i.e. milestones) of the search journey, while still taking into account the commonality of guest preference throughout all stages of the search journey.

Furthermore, the mixture of positive (e.g. booking) and negative (e.g. cancellation) stages poses additional challenges. Specifically, the negative stages can be initiated from either guest or host, with varying correlations to other milestones in the same search journey. Explaining how the ranking model balances these often conflicting stages is an important consideration in a real world product.

Finally, the high level concept of “Guest Journey” applies beyond the Airbnb stays ranking (e.g., experience ranking, email marketing, etc). The main differences are in terms of product-specific milestones (e.g. negative stages such as unsubscription in email marketing). Crafting a model architecture that optimizes guest journey in Airbnb stays ranking, while also easily applied to other product use cases, is strategic to the success of the Airbnb ecosystem. 

In this paper, we present \textbf{Journey Ranker}, a general multi-task model architecture with the following main contributions:
\begin{itemize}
    \item \textbf{Learning both Positive and Negative  Milestones:} Journey Ranker leverages intermediate guest actions to help the guest progress towards positive milestones, and avoid negative milestones throughout the guest journey.
    \item \textbf{Balancing Guest Journey:} Journey Ranker introduces a Combination module (Section \ref{section:combination_module}) to balance the positive and negative milestones across the guest journey.
    \item \textbf{Modular and Extensible Model Architecture:} Journey Ranker is designed to consist of four modules with clear separation of concerns. This modular design allows it to be easily extended to other use cases beyond Airbnb stays ranking, which will be discussed in Section \ref{section:online_experiments_other_search}.
\end{itemize}

In this paper, we focus our presentation on a particular instantiation of the Journey Ranker for the Airbnb Stays ranking. Nevertheless, Journey Ranker is applicable to other search use cases where the user decision journey is long (e.g. e-commerce sites, real-estate listing sites, dating sites, etc), by choosing the appropriate milestones in the Base Module described in ~\ref{section:base_module}. Journey Ranker is also applicable when there are negative business outcomes (e.g. customer support tickets, poor product rating, product returns, etc) to be avoided, by modeling them in the Twiddler Module described in ~\ref{section:twiddler_module} and then balanced by the Combination Module described in ~\ref{section:combination_module}. Journey Ranker could also be applied to other non-search journeys (e.g. email marketing, customer acquisition, etc) by configuring different positive and negative milestones. We present our successes in deployment in other use cases in Section ~\ref{section:online_experiments_other_search}.

The rest of the paper is organized as follows: In Section \ref{section:related_work}, we describe related work on search and recommendation systems. In Section \ref{section:airbnb_ranking}, we present the Airbnb Stays Ranking problem formulation and the baseline. Next, we present Journey Ranker in Section \ref{section:journey_ranker} and the offline and online experiment results in Section \ref{section:experiment_results}. Finally, we discuss possible future directions in Section \ref{section:conclusion}.

\section{Related Work}
\label{section:related_work}
The majority of previous works in Search and Recommendation involves training deep learning models to optimize for a single business objective such as app installs in Mobile App Store \cite{wide_deep_2016}, watch time in video recommendation \cite{youtube_2016}, pCTR in advertisement \cite{DCN_v1}, bookings of vacation rentals \cite{Haldar_2019}, etc. To deal with multiple potentially conflicting business objectives, one approach is to build an ecosystem of stand-alone models for each business objective \cite{Haldar_2019}, and apply them as "second-stage re-rankers" to finetune the top K results after the main ranker. In contrast, Journey Ranker incorporates all the different guest actions corresponding to various objectives in a single multi-task model trained end-to-end \cite{Caruana2004MultitaskL}.

Recently, multi-task modeling has been an active topic in many Search and Recommendation use cases such as video recommendation \cite{Youtube_2019}, feed ads \cite{GoogleFeedAds_2022}, etc., when there are multiple business metrics. Leveraging a multi-task approach allows sharing of feature representation, resulting in better performance for each task through transfer learning, while achieving lower overall serving cost. In contrast, for Airbnb Stays ranking, the main objective is uncancelled booking. Improving other metrics such as clicks without moving uncancelled booking implies a degraded ease of search experience for guests. Thus, our application of multi-task modeling is different from standard multi-objective settings in the sense that we are not trying to improve each of the objectives through shared representation. Rather, we leverage multi-task approach to let the model have a more comprehensive view of the guest search journey when optimizing for the uncancelled booking objective.

Such a strategy is similar to previous industry use cases that leverage multi-task modeling to model multi-step conversions, such as in e-commerce \cite{ESMM_2018} and in customer acquisition \cite{Xi2021ModelingTS}. Journey Ranker is inspired by these previous works, especially \cite{ESMM_2018} in our design of Base Module in Section \ref{section:base_module}. In addition, we also extend previous work by adding new modules such as the Twiddler Module to model other negative search milestones and the Combination Module to better balance conflicting milestones throughout the guest search journey.

\begin{figure}
  \centering
  \begin{minipage}[b]{0.5\linewidth}
      \centering
      \includegraphics[width=\linewidth]{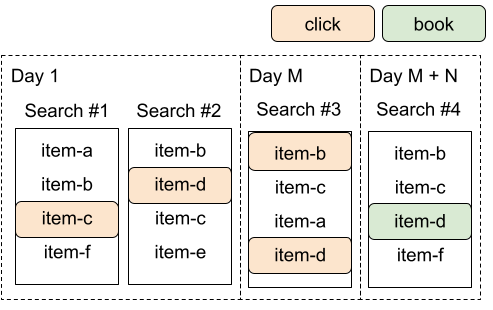}
      \Description{A guest makes four searches across three different days, and clicked a few items before finally booking an item.}
      \subcaption{Example search journey}
  \end{minipage}%
  \begin{minipage}[b]{0.5\linewidth}
      \centering
      \includegraphics[width=\linewidth]{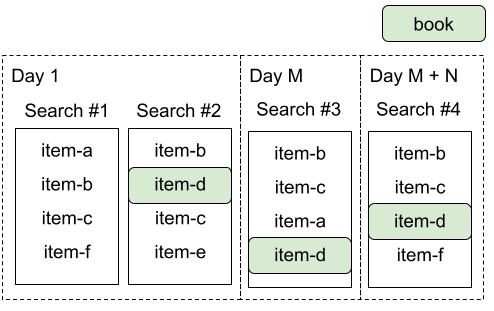}
      \Description{The booked item is marked with booking label across all the three searches that contains it}
      \subcaption{Example label attribution}
  \end{minipage}
  \caption{An example Airbnb guest search journey and its label attribution.}
  \label{figure:example_user_journey}
\end{figure}

\section{Airbnb Stays Ranking}
\label{section:airbnb_ranking}

\subsection{Notations}
\label{section:shorthand_notation}
To ease presentation, here are the notations we will use throughout the paper to describe an Airbnb user action, which will be used to denote an output, or a loss.

\begin{itemize}
    \item \textit{imp} - An impression of a search result in a search.
    \item \textit{c} - A click of a search impression.
    \item \textit{lc} - A long click of a search impression, defined by engagement of the listing details page after a click.
    \item \textit{pp} - A click to open the payment page for a listing.
    \item \textit{req} - A reservation request for a listing from a guest.
    \item \textit{book} - An acceptance (by host) of a reservation request.
    \item \textit{unc} - An uncancelled booking of a listing.
    \item \textit{rej} - A rejection of a reservation request.
    \item \textit{cbh} - A cancellation by host.
    \item \textit{cbg} - A cancellation by guest.
\end{itemize}

We use \textit{U}ppercase to denote a Set, \textit{l}owercase to denote an instance of a Set, and subscript to denote relationship between two sets, for example:

\begin{itemize}
    \item \textit{U} is a set of guests.
    \item \textit{u} is a guest within \textit{U}. 
    \item \begin{math} S_u \end{math} is a set of searches for guest \textit{u}.
\end{itemize}

Finally, we use \begin{math} y_t \end{math}, and \begin{math} Loss_{t} \end{math} to indicate the output, and loss for task (or module) \textit{t} respectively.

\subsection{Previous Baseline}
\label{section:baseline_model}
Figure \ref{figure:example_user_journey} (left diagram) shows a simplified example of a typical guest search journey, where the guest makes multiple searches over multiple days and clicks on multiple search results to view their detail pages, before finally making a reservation request for a listing. The reservation request is then instantly-accepted if it is an Instant Book listing, otherwise the host manually decides whether to accept or reject the reservation request. Finally, both the guest and host may also cancel the booking before check-in. Thus, an \textbf{Uncancelled Booking} is one of the main metrics we use to evaluate the success of a guest search journey.

During model training of previous baseline, we label all of the search impressions of the booked listing in all of the searches the guest made before the booking occurred as "positive labels" as shown in Figure \ref{figure:example_user_journey} (right diagram). The goal of the Airbnb Stays Ranking model is then formulated as ranking the positive labels higher than all the other search impressions; and we do this for all guest searches across the search journey so that guests can find their ideal listing as early as possible.

Specifically, we want to learn model parameters \begin{math} \theta \end{math} to minimize the loss across the searches, \begin{math} S_{u} \end{math}, for all guests, \textit{U}.

\begin{equation}
  \argmin_{\theta}{\sum_{u \in U}{\sum_{s \in S_u}{Loss_{unc}{(s | \theta)}}}}
  \label{equation:ranking_loss}
\end{equation}

Note that the \begin{math} Loss_{unc} \end{math} is defined per search with the goal of ranking the search impressions associated with the booking higher than every other search impressions. The loss could be either listwise, or pairwise loss using standard learning-to-rank approaches \cite{burges2010ranknet} \cite{ltr2007}.

The previous production Airbnb Stays Ranking model then generates a score for each listing for ranking based on the model parameters \begin{math} \theta \end{math}, listing features \begin{math} F_L \end{math}, and context (i.e query and guest) features \begin{math} F_C \end{math} using a deep neural network consisting of two towers of fully connected layers with non-linear activation units \cite{Haldar_2020} to predict uncancelled booking.

\subsection{Disadvantages of Previous Baseline}
\label{section:beyond_conversion}
While optimizing for booking aligns well with our business metric, there are a few disadvantages that are worth noting:
\begin{itemize}
    \item We ignore searches from bookers without the booked listings, e.g. Search \#1 in Figure \ref{figure:example_user_journey} (right diagram).
    
    \item We equally treat all non-booked listings as negative examples, but in reality, some of them have clicks while others don't, e.g. \textit{item-b} in Search \#3 in Figure \ref{figure:example_user_journey} (left diagram).
    
    \item We ignore all searches from non-bookers, including those guests who had reservation requests but were rejected by the hosts, which could provide useful insights for understanding guest and host preferences.
\end{itemize}

Given that both non-bookers and non-booking guest actions (eg. search clicks) are each one and two orders of magnitude bigger than bookers and booking events respectively, we are potentially missing out a significant amount of data to better extract guest and host preferences. Furthermore, we are also subject to \textit{sample selection bias} problem \cite{samplebias2004} as we are serving a ranking model learned only from our past bookers to all our guests.

One important consideration for Airbnb is that conversion remains the ultimate goal. Specifically, we don't want a model that increases the number of clicks without increasing bookings, as that will imply that each booking involves more effort across a longer search journey to find the ideal listing.

Our previous attempts to remedy these issues included building a single multi-task model with booking and long click \cite{Haldar_2019}, and building multiple single-task models that are then combined online during serving using weights found through grid search. The former attempt increased long clicks but failed to move bookings. The latter attempt successfully increased bookings but had poor stability, resulting in out-of-date weights and a decline on on bookings a few months after the original launch.

In the next section, we present \textbf{Journey Ranker}, a novel multi-task model architecture trained end-to-end. Journey Ranker successfully drove significant business metric improvements without the downside of our previous approaches mentioned above.


\section{Journey Ranker for Stays Ranking}
\label{section:journey_ranker}

\begin{figure}
  \centering
  \includegraphics[width=.8\linewidth]{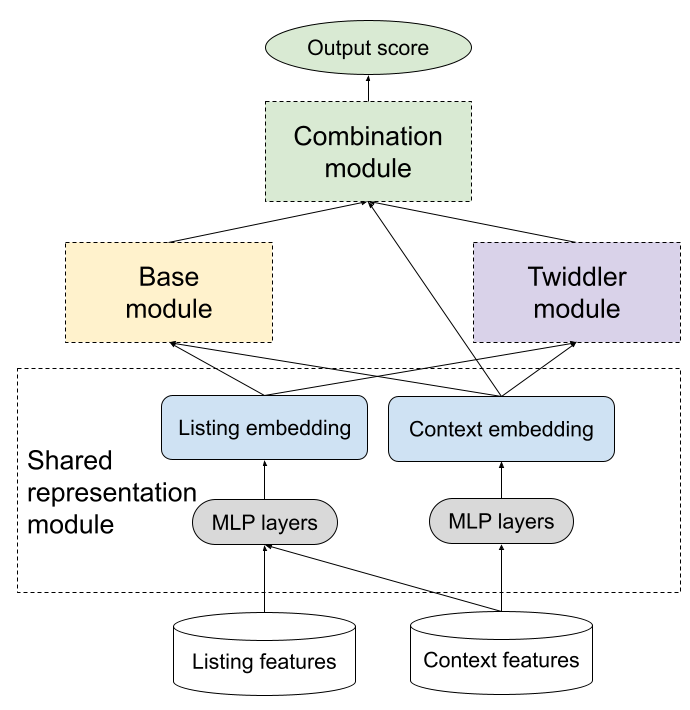}
  \Description{A simplified view of the Journey Ranker model architecture, consisting of multiple modules - shared representation module, base module, twiddler module, and combination module.}
  \caption{A simplified view of the Journey Ranker model architecture, consisting of multiple modules - Shared Representation Module, Base Module, Twiddler module, and Combination module.}
  \label{figure:simplified_architecture}
\end{figure}

Journey Ranker architecture consists of multiple modules trained end-to-end using multi-task techniques, and a simplified view of the architecture is shown in Figure \ref{figure:simplified_architecture} for the instantiation for our Airbnb Stays Ranking use case. The input to Journey Ranker are listing features \begin{math} F_L \end{math}, and context (i.e query and guest) features \begin{math} F_C \end{math}. Both \begin{math} F_L \end{math} and \begin{math} F_C \end{math} consist of both continuous and categorical features, and they are transformed accordingly through either normalization or embedding projection as described in previous work \cite{Haldar_2019}.

The Shared Representation Module takes both \begin{math} F_L \end{math} and \begin{math} F_C \end{math} as input and feeds them through multi-layer perceptrons (MLP) with non-linear activation units to generate listing embedding, \begin{math} Emb_L \end{math} and context embedding, \begin{math} Emb_C \end{math}. Finally, \begin{math} Emb_L \end{math}  and \begin{math} Emb_C \end{math} are fed to the three remaining modules (Base Module, Twiddler Module, and Combination Module) to generate the final output score. These modules will be presented in detail in the following subsections.

One thing to note is that if we remove the Twiddler Module and the Combination Module, and reduce the Base Module to be a single-task model optimizing for just uncancelled booking, then the output of the Base Module is equivalent to our previous production model architecture described in Section \ref{section:baseline_model}.

\subsection{Base Module: Reaching Positive Milestones}
\label{section:base_module}

\begin{figure*}
  \centering
  \includegraphics[width=.8\linewidth]{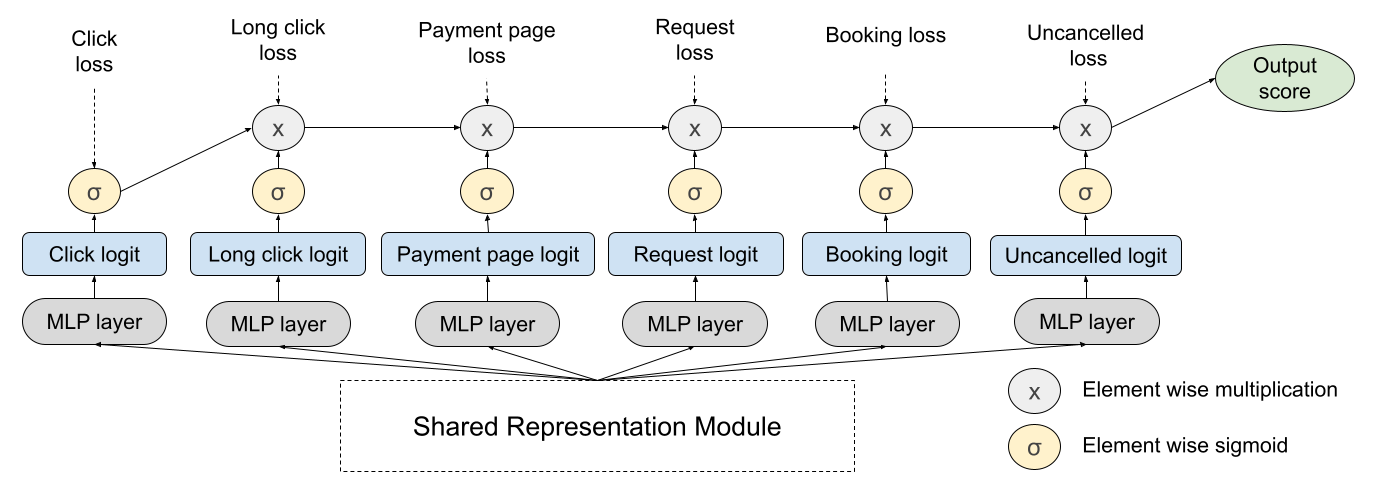}
  \Description{Model architecture for the Base Module consisting of six sequential guest actions of clicks, long clicks, payment page views, reservation requests, bookings, and uncancelled bookings.}
  \caption{Airbnb Stays Ranking's Instantiation of Base Module.}
  \label{figure:base_module_architecture}
\end{figure*}

The responsibility of the Base Module is to produce a single score per search result that is optimized for the final positive milestone (e.g. uncancelled booking for Airbnb stays ranking). As described in Section \ref{section:beyond_conversion}, we withhold a lot of information from the model if we simplify the complex Airbnb Stays ranking use case into a binary formulation of uncancelled booking vs search impression. 

Thus, in Journey Ranker for Airbnb Stays Ranking, we reformulate the ML problem formulation for Base Module into helping guests reach each of the positive search milestones (i.e. improving the conversion funnel), culminating in an uncancelled booking. Motivated by \cite{ESMM_2018}, we also model the six positive guest search milestones as sequential guest actions that are linked together using the chain rule of probability. Specifically, we decompose the probability of uncancelled booking into each of the six guest milestones, i.e. Click \textit{c}, Long Click \textit{lc}, Payment page \textit{pp}, Reservation request from guest \textit{req}, Acceptance of reservation request by host \textit{book}, and Uncancelled Booking \textit{unc}, with the following equations:

\begin{flalign}
\begin{split}
P(unc) =&\ P(unc\ |\ book) * P(book) \\ 
P(book) =&\ P(book\ |\ req)\ * P(req) \\
P(req) =&\ P(req\ |\ pp) * P(pp) \\
P(pp) =&\ P(pp\ |\ lc) * P(lc) \\
P(lc) =&\ P(lc\ |\ c) * P(c)
\end{split}
\label{equation:base_equation}
\end{flalign}

where each of the terms on the left hand side of the equation is actually a joint probability of all its preceding guest milestones, omitted for brevity. For example, \begin{math} P(unc) \end{math} is a joint probability, i.e. \begin{math} P(unc \cap book \cap req \cap pp \cap lc \cap c\ | \ imp) \end{math}.

The instantiation of the Base Module for Airbnb Stays Ranking is in Figure \ref{figure:base_module_architecture}, where for each of the guest milestones, we generate a corresponding task logit (i.e. blue box) by running a multi-layer perceptron (MLP) on top of the Listing Embedding \begin{math} Emb_L \end{math} and Context Embedding \begin{math} Emb_C \end{math} from the Shared Representation Module. In other words, the six blue boxes in Figure \ref{figure:base_module_architecture} correspond to the six conditional probabilities defined in Equation \ref{equation:base_equation}.

To train the Base Module for the Journey Ranker, we apply multiple losses, one for each of the joint probabilities in Equation \ref{equation:base_equation} as shown in the Figure \ref{figure:base_module_architecture}. For example, for a specific task such as payment page, all the search impressions that led to a payment page view are considered positives and the rest are considered negatives. 


The final loss for the Base Module is the sum of each of the losses from each individual tasks:

\begin{equation}
    Loss_{Base} = \sum_{task} Loss_{task}
    \label{equation:base_module_loss}
\end{equation}

where, using the same notation as Equation \ref{equation:ranking_loss}, \begin{math} Loss_{task} \end{math} is:

\begin{equation}
    Loss_{task} = \sum_{u \in U} \sum_{s \in S_u} loss_{task}(s | \theta) * w_{task}
    \label{equation:base_task_loss}
\end{equation}

where \begin{math} loss_{task}(s | \theta) \end{math} corresponds to loss for a given search \textit{s} computed using standard learning-to-rank approaches \cite{burges2010ranknet} \cite{ltr2007}, and \begin{math} w_{task} \end{math} is a normalization term for a given task to avoid tasks with a lot of occurrences (e.g. clicks) from dominating the final loss. 

E.g., suppose for the long click task, we use softmax listwise loss for \begin{math} loss_{lc}(s | \theta) \end{math}, where each search that consists of N long clicks will result in N softmax loss. In that case, \begin{math} w_{lc} \end{math} will be the empirical fraction of long clicks that will convert into an uncancelled booking in our training data. The intention behind designing such normalization term is:
\begin{itemize}
    \item Each positive milestone is assigned credit proportional to its likelihood to convert into the final uncancelled booking.
    \item The Base Module is learning multiple different but related reformulations of an uncancelled booking (i.e. through intermediate search milestones), which could help the model learn a better Shared Representation Module.
\end{itemize}

Overall, the design of Base Module as described above allows us to better capture the sparse, long-duration, exploratory, and assorted guest activities at Airbnb, which is an important challenge for the Airbnb product use case, as described in the Section \ref{section:introduction}. Note that other use cases could have different number of positive milestones corresponding to their specific guest journey.

\subsection{Twiddler Module: Avoiding Negative Milestones}
\label{section:twiddler_module}

\begin{figure}
  \centering
  \includegraphics[width=.8\linewidth]{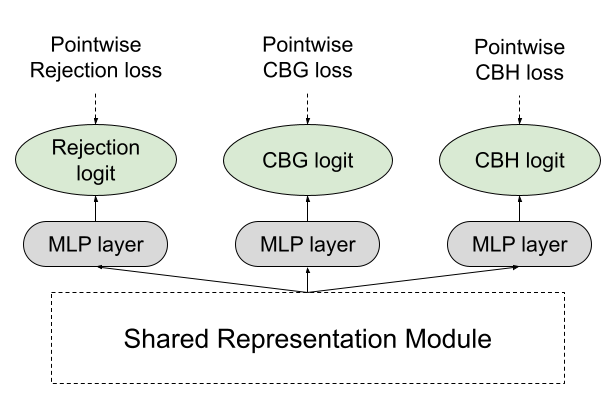}
  \Description{Airbnb Stays Ranking's Instantiation for the Twiddler Module consisting of three negative milestones of rejections (of reservation requests by host), cancellation by guest (CBG), and cancellation by host (CBH).}
  \caption{Airbnb Stays Ranking's Instantiation for the Twiddler Module.}
  \label{figure:twiddler_module_architecture}
\end{figure}

The responsibility of the Twiddler Module is to produce a score per search result for each of the negative milestones. The instantiation of Twiddler Module for Airbnb Stays Ranking can be found in Figure \ref{figure:twiddler_module_architecture}, where we have three Twiddler tasks, i.e. rejection, cancellation by host, and cancellation by guest. Twiddler Module is learned in the same multi-task setting with each of the Twiddler tasks receiving as input the embeddings from the Shared Representation Module, and passing the embeddings through a multi-layer perceptron (MLP) to generate the final logit. Similar to Equation \ref{equation:base_module_loss}, the loss for Twiddler Module is just the sum of each of the twiddler losses:

\begin{equation}
    Loss_{Twiddler} = \sum_{task} Loss_{task}
    \label{equation:twiddler_module_loss}
\end{equation}

where \begin{math} Loss_{task} \end{math} is a binary classification loss for each of the negative milestones (eg. rejected reservation request vs accepted reservation request for the Rejection Twiddler task).

One might wonder why we can't reuse the corresponding conditional probabilities (eg. \begin{math} P(book\ |\ req) \end{math}) in Equation \ref{equation:base_equation} from the Base Module in Section \ref{section:base_module} for the twiddlers here. The reasons are:

\begin{itemize}
    \item In Base Module, we apply the losses directly on the joint probabilities instead of the conditional probabilities. Thus, the conditional probabilities are biased and not accurate \cite{escm2022}.
    \item We achieve more accurate modeling of cancellations by separately modeling cancellations initiated by the guest and host. However, since the Base Module currently uses a linear multi-step conversion setup, it cannot implement this separation (i.e. requires a "forked" conversion funnel instead of a linear conversion funnel).
    \item These negative search milestones are rarer (<1\% to ~10\% depending on the action), leading to class imbalance. They are also delayed outcomes where the final label could arrive after a long time has elapsed. By designing them as its own separate module, we can optimize each of the twiddlers further to account for its unique circumstances such as different sampling strategies, etc. in future works.
\end{itemize}

Overall, the Twiddler Module allows us to incorporate negative milestones from both guest and host in the search journey, which is one of the unique challenges for Airbnb product described in Section \ref{section:introduction}. Note that other use cases would have different twiddlers, corresponding to different use case specific negative milestones.

\subsection{Combination Module: Balancing Guest Journey}
\label{section:combination_module}

\begin{figure*}
  \centering
  \includegraphics[width=.8\linewidth]{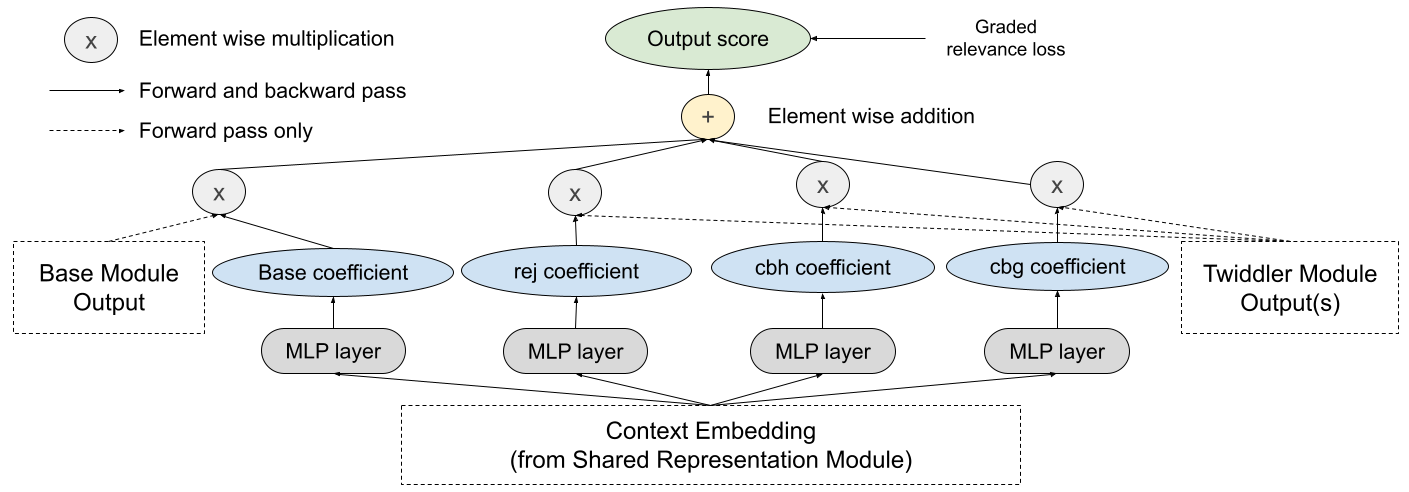}
  \Description{Model architecture for the Combination Module.}
  \caption{Airbnb Stays Ranking's Instantiation for the Combination Module.}
  \label{figure:combination_module_architecture}
  \vspace{-2mm}
\end{figure*}

Finally, the responsibility of the Combination Module is to balance the output from the Base Module that is optimized for positive milestones and the outputs from the Twiddler Module that are optimized for negative milestones. Figure \ref{figure:combination_module_architecture} shows the Airbnb Stays Ranking's instantiation for the Combination Module, which can be represented as the following:

\begin{flalign}
\begin{split}
    y_{Combination} = y_{Base} * \alpha_{Base}
      + \sum_{t \in twiddler} y_t * \alpha_t
    \label{equation:combination_equation}
\end{split}
\end{flalign}

where the final output score, \begin{math} y_{Combination} \end{math}, is obtained by applying a linear combination on top of the Base Module score, \begin{math} y_{Base} \end{math}, and the Twiddler Module scores, \begin{math} y_t \end{math}. Note that each coefficient \begin{math} \alpha_x \end{math} in the linear combination is a scalar learned end-to-end in the Combination Module, generated using a MLP that takes in the Context Embedding \begin{math} Emb_C \end{math} from the Shared Representation Module. 

There were several key design choices made, i.e.:

\begin{itemize}
    \item \textbf{A context-dependent coefficient $\alpha$ instead of a global coefficient:} As shown in Figure \ref{figure:airbnb_journey_flowchart}, the Airbnb search journey is very long and the needs of the guest change as they gather more information along the journey. E.g., if a guest already experienced a negative search milestone in the journey, we want to emphasize avoiding another negative search milestone to prevent churn. A context-dependent coefficient allows the model to learn to balance the guest search journey better as we know more about the guest.
    \item \textbf{A linear combination instead of a non-linear combination of the various outputs:} We attempted both options and they perform similarly in the offline evaluation. Thus, we opted for the simpler linear combination manner as it allows better product interpretability, which will be presented in more details in Section \ref{section:decoding_combination_module}.
    \item \textbf{Freezing the gradient update from Combination Module to the Base and Twiddler Module:} Note that in the Figure \ref{figure:combination_module_architecture}, we do not have backward pass from the Combination Module to the output from the Base Module and Twiddler Module. We designed it this way so that the Combination Module can focus merely on learning how to combine these outputs without changing their semantics.
\end{itemize}

Such designs allow us to better understand what the module is learning, which is explored further in Section \ref{section:decoding_combination_module}.

Finally, we devise an objective function for the Combination Module so that it learns to properly balance positive and negative search milestones. To do that, we construct a pairwise loss for the final output of the Combination Module to learn graded relevance of different guest actions based on the following pairwise relationship:
\begin{equation}
    unc > click > imp > cbg\ /\ cbh\ /\ rej
    \label{equation:combination_graded_relevance}
\end{equation}

Specifically, uncancelled booking is considered the best outcome, followed by clicks and then all of the unclicked search impressions. All the negative search milestones (i.e. cancellation by guest, cancellation by host, and rejection) are considered the worst outcomes, even worse than the unclicked search impressions. 

Finally, the full loss is the sum of losses across all modules, i.e.
\begin{equation}
    Loss_{Total} = Loss_{Base} + Loss_{Twiddler} + Loss_{Combination}
    \label{equation:journey_ranker_full_loss}
\end{equation}

Note that other use cases could define Equation ~\ref{equation:combination_graded_relevance} differently to optimize for their own version of a satisfied user (e.g. good product rating without product returns in e-commerce sites).

\subsection{Richer Training Data}
To enable Journey Ranker to learn from various different guest actions, we make the following changes to the training data:
\begin{itemize}
    \item \textbf{Labeling of Search Impressions:} We label each search impression with all the labels that are applicable (i.e. multi-label) - both positive guest actions (i.e. click, long click, payment page, reservation request, accepted booking, and uncancelled booking) and negative guest actions (i.e. rejection, cancellation by guest, and cancellation by host).
    \item \textbf{Number of Searches:} By leveraging guest actions such as clicks, we can now leverage two orders of magnitude more data for our model training. However, this introduces both scalability complexity and more noise to our model training due to lower intent exploratory guests. In the end, we included only all the searches that led to a payment page view after evaluating empirically the trade off between number of searches and the model performance. This leads to just around 50\% more searches compared to before.
\end{itemize}

\section{Experimental Result}
\label{section:experiment_results}
For offline evaluation, we use Normalized Discounted Cumulative Gain (NDCG) with binary relevance score, where uncancelled booking is assigned a relevance score of 1 and all other search impressions are assigned a relevance score of 0.

For model training, we use around 500 millions searches for training, and all the models are trained using Tensorflow. All the models are trained with five different initializations and we report the average offline performance along with its 95\% confidence interval. Both the training throughput and the end-to-end model latency are similar compared to the previous baseline, as we are only adding +9.2\% parameters.

\subsection{End-to-end Offline Evaluation}
\begin{table}[]
\begin{tabular}{|c|c|c|ll}
\cline{1-3}
\textbf{Model} & \textbf{Offline NDCG} & \multicolumn{1}{l|}{\textbf{\# Parameters}} &  &  \\ \cline{1-3}
Baseline       & +0.0\% ($\pm$0.03\%)      & +0.0\%                                        &  &  \\ \cline{1-3}
Journey Ranker & +0.48\% ($\pm$0.05\%)  & +9.2\%                                      &  &  \\ \cline{1-3} 
\end{tabular}
    \caption{End-to-end offline evaluation for the Airbnb Stays' instantiation of Journey Ranker described in Section \ref{section:journey_ranker}.}
    \label{table:overall_offline_evaluation}
    \vspace{-2mm}
\end{table}

Table \ref{table:overall_offline_evaluation} shows the offline evaluation of the Journey Ranker described in Section \ref{section:journey_ranker} compared to the previous Baseline model described in Section \ref{section:baseline_model}. Overall, Journey Ranker has a statistically significant improvement of +0.48\% NDCG compared to the baseline, with a small 9.2\% increase in number of trainable parameters. The offline result is encouraging because for Airbnb Stays ranking use case:
\begin{itemize}
    \item An offline evaluation of +0.3\% is considered promising to land a statistically significant booking gain in online A/B test, which we will present in Section \ref{section:online_experiments} later.
    \item For the baseline Airbnb Stays Ranking, a +50\% in trainable parameters (through making the network deeper or wider) will only have about 0.05-0.1\% offline NDCG gain.
    \item Despite having a more complicated model architecture (10 tasks!), we only incur a small 9.2\% increase in number of trainable parameters because each of the new task-specific networks is a small (single-layer) network that shares the same representation layers before it. This means we still have more opportunity to improve gain further by increasing the capacity for each task using other techniques \cite{Meyerson2018PseudotaskAF} \cite{smallhead_2022}.
\end{itemize}

In the following subsections, we will examine in more detail how each of the Modules within Journey
Ranker behaves to shed some light on where the improvements come from.

\subsection{Discussion of Alternative Designs}
\label{section:alternative_designs}
In this section, we describe some of the alternative designs that we attempted and present our findings.

\textbf{A single end-to-end module} - Could we design a single task model that directly optimizes for graded relevance described in the Equation \ref{equation:combination_graded_relevance}? We attempted this and the naive application of the same graded relevance resulted in > 1\% drop in offline NDCG. We can get a better result (but still worse than Journey Ranker's) if we tune the relative weights between each of the labels in the graded relevance setup. We hypothesize that by giving the model full freedom, the different labels will interfere with each other to compete for the same model parameters, causing the more prevalent tasks (eg. click) to dominate the final output. By designing the model architecture using well-defined modules and tasks with separation of concerns, each of the tasks will have its own task-specific parameters to learn knowledge specific to itself, while contributing to the shared representation layer for the common knowledge.

\textbf{Non-uniform loss weights} - This motivates another question of whether we should tune the relative weights of losses for each of the tasks in Journey Ranker? Note that, as described in Equation \ref{equation:journey_ranker_full_loss}, Journey Ranker currently uses a simple unweighted sum of losses across each of the modules (and thus each of the tasks). Interestingly, we tried to manually tune these weights but did not achieve very statistically significant offline results that could justify the increase in complexity. However, we believe there is opportunity in the future to revisit using an auto-hyperparameter tuning service or other surrogate tricks \cite{GoogleFeedAds_2022} to improve this further.

\textbf{Dealing with negative transfer} - We also tried various previous works on dealing with gradient interference \cite{gradnorm} and multi-task loss weighting \cite{kendall_uncertainty} for Journey Ranker and did not see stat-sig improvement in offline NDCG. We hypothesize that this is because all of the ten tasks we leverage in Journey Ranker are ten different aspects of understanding the same guest preference (i.e. uncancelled booking), and thus don't suffer from standard multi-task challenges where the tasks are very different from each other.

\subsection{Can we simplify the Base Module?}
\label{section:base_module_eval}

\begin{table}[]
\begin{tabular}{|c|c|c|c|l}
\cline{1-4}
\textbf{Tasks Setting}   & \textbf{Offline NDCG}                                          & \multicolumn{1}{l|}{\textbf{Params}} & \multicolumn{1}{l|}{\textbf{Searches}} &  \\ \cline{1-4}
Baseline (Unc)                  & +0\% ($\pm$0.03\%)    & +0.0\%                                  & +0\%                                      &  \\ \cline{1-4}
Req + Book + Unc & +0.06\% ($\pm$0.03\%) & +1.8\%                                  & +25\%                                     &  \\ \cline{1-4}
Click + Unc          & +0.22\% ($\pm$0.02\%)\ & +1.8\%                                  & +50\%                                     &  \\ \cline{1-4}
All 6 Tasks                  & +0.31\% ($\pm$0.02\%)  & +6.0\%                                  & +50\%                                     &  \\ \cline{1-4}
\end{tabular}
\caption{Ablation study on the tasks used in the Base Module for Airbnb Stays Ranking, without Twiddler and Combination Module. We present four settings that have both practical product semantics in Airbnb Stays Ranking and also stat-sig difference in offline NDCG across each pairs.}
\label{table:base_module_evaluation}
    \vspace{-6mm}
\end{table}

In the Base Module, we have six tasks (clicks, long clicks, payment page, requests, accepted bookings, and uncancelled bookings) corresponding to six different search milestones in the guest search journey. One might wonder whether we need all six of these? 

Table \ref{table:base_module_evaluation} shows the ablation study of the tasks used in the Base Module, without the Twiddler and Combination Module to better understand the effect. We presented four choices of tasks that have useful product semantics in Airbnb and also statistically significant improvement in offline NDCG across each pair in the setup. There are a few useful learnings:
\begin{itemize}
    \item \textbf{Quantity of searches:} As expected, as the number of searches used for training the Base Module increases, the offline NDCG improves (i.e. from row 1 to 3 in Table \ref{table:base_module_evaluation}).
    \item \textbf{"Diversity" of searches:} Adding new searches that are more different from existing searches could lead to more gain. E.g., in Table \ref{table:base_module_evaluation}, adding +25\% more searches with reservation requests (that was ultimately rejected or cancelled) only had a +0.06\% NDCG improvement (row 2 vs row 1). In contrast, if we add +20\% more searches with clickers (that didn't have any reservation requests), we had a +0.16\% NDCG improvement (row 3 vs row 2). This aligns with our hypothesis that by only utilizing searches from uncancelled bookers, we might not generalize well to the non-bookers, who might have very different guest behavior.
    \item \textbf{Usefulness of intermediate search milestones:} In table \ref{table:base_module_evaluation}, if we compare row 4 vs row 3, we have a +0.09\% offline NDCG improvement despite having the same number of searches. This suggests that adding more intermediate search milestone labels to better differentiate the guest preference on each of the search impressions is useful. In the future, we plan to try other efforts to maximize the signals derived from each search, such as through different model architectures for Base Module \cite{gmcm2020} \cite{esm2_2020}.
\end{itemize}

In addition to offline NDCG for the uncancelled booking as the label, we also look at the offline NDCG for all the other labels such as clicks, long clicks, payment pages, and requests. We omit the detailed result for brevity, and the high level observation is that the offline NDCG for each of these search milestones also has a statistically significant gain. This gives another insight on where the offline NDCG gain for uncancelled booking comes from - by improving conversion for each of the intermediate search milestones, the final uncancelled booking will also naturally increase.

\subsection{What does the Combination Module learn?}
\label{section:decoding_combination_module}

\begin{figure}
  \centering
  \includegraphics[width=.8\linewidth]{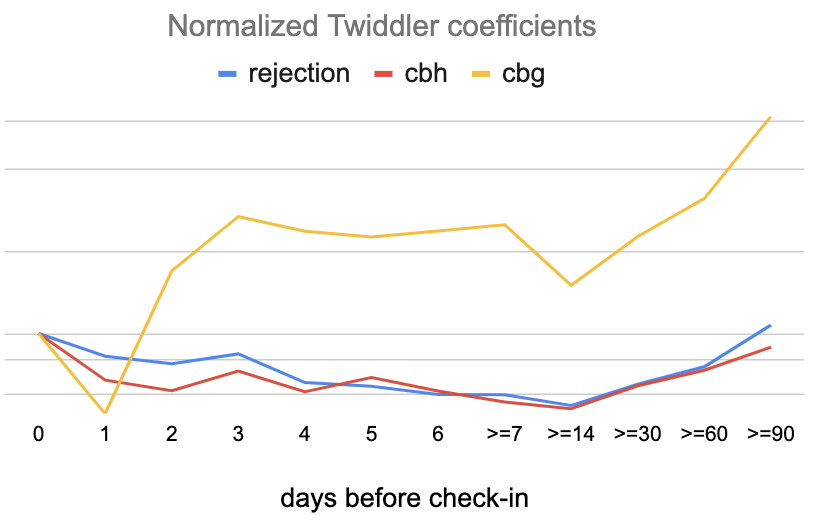}
  \Description{For rejection and cancellation by host, the coefficients have a U-shaped while for the cancellation by guest, the coefficients is increasing.}
  \caption{NTC when segmented by days ahead of the check-in for a search query. Each NTC curve is normalized by NTC when days before check-in = 0 for ease of visualization.}
  \label{figure:days_ahead_combination_coeff}
\end{figure}

\begin{figure}
  \centering
  \includegraphics[width=.8\linewidth]{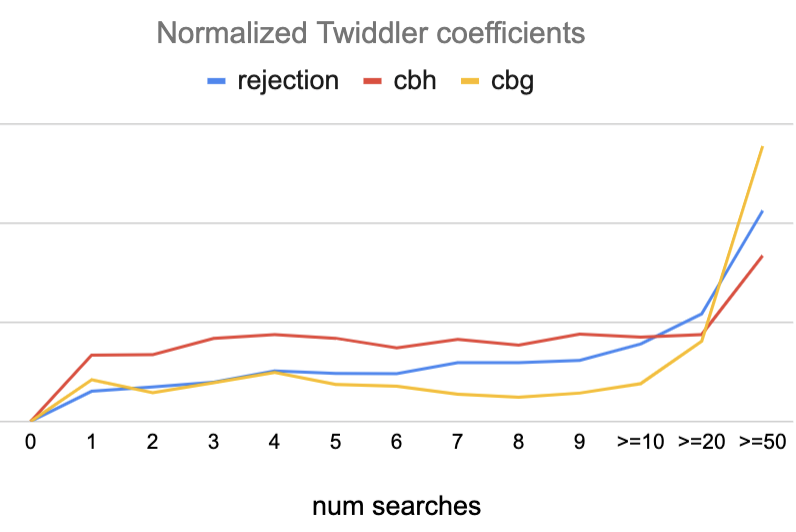}
  \Description{As guest progresses further in the search journey with more number of previous searches, the twiddler coefficients increase.}
  \caption{NTC when segmented by guest's number of previous searches. Each NTC curve is normalized by NTC when num searches = 0 for ease of visualization.}
  \label{figure:num_searches_combination_coeff}
  \vspace{-4mm}
\end{figure}

As described in Equation \ref{equation:combination_equation}, the output of the Combination Module is a linear combination of the outputs from the Base Module and the Twiddler Module. Recall that each coefficient is learned end-to-end based on the Context Embedding \begin{math} Emb_C \end{math}, and thus are dependent on Context Features. This allows us to study how these coefficients change given different guest or query features to try to decipher what the Combination Module learns. 

\textbf{Definition} - We compute the "\textit{Normalized Twiddler coefficient}" (NTC), which normalizes the coefficient for each of the Twiddler outputs, $\alpha_{twiddler}$ against the coefficient of the Base Module, $\alpha_{base}$. We can now plot a diagram for each context feature, where the x-axis is the various values for a particular feature, and the y-axis is the NTC metric to understand how the NTC changes as the feature value changes. Note that a higher NTC indicates that the Combination Module is giving more importance to a particular Twiddler task, i.e. more importance to avoiding a specific negative search milestone, compared to the Base Module.

Next, we present examples of what the Combination Module learns with respect to two context features:

\textbf{An example query feature} - Figure \ref{figure:days_ahead_combination_coeff} shows how NTC moves with respect to the query feature "days ahead of check-in". Here, we notice that there are two different shapes for the NTCs:
\begin{itemize}
    \item Rejection and cancellation by host are both negative search milestones initiated by the host, and thus it makes sense that both of their NTC curves are similar - i.e. a U-shaped curve. This shape aligns with our product intuitions for when the search product should try to minimize rejections or cancellation by hosts with respect to this query feature. Specifically, if the trip is close to the query time, there is not enough time for guests to rebook. On the other hand, if the trip is far from the query time, minimizing the likelihood of future unforeseen circumstances that cause the host to to reject / cancel before the trip is crucial.
    \item We hypothesize that the NTC curve for cancellation by guest is increasing for trips that are further out to prevent the guest from cancelling due to buyer’s remorse (eg. found a better listing or realizing that the listing does not satisfy their need due to location or missing amenities etc).
\end{itemize}

We hypothesize that the model learns these NTC curves through a mixture of reasons, such as:
\begin{itemize}
    \item Differences in prevalence of the negative search milestones for each query segment.
    \item Label attribution in training data - E.g., if the trip is coming up soon, the guest might not have time to rebook on Airbnb or might have opted for other options such as a traditional hotel. On the other hand, if the trip is far into the future, the guest still has plenty of time to rebook. Since we use a max window of K days to define a guest search journey, the successive searches might be grouped into a separate search journey. Regardless of the cases, the original search journey is now considered a failed journey, and the model learns to try to avoid these negative outcomes.
\end{itemize}

\textbf{An example guest feature} - Figure \ref{figure:num_searches_combination_coeff} shows how NTC varies against the guest's number of previous searches. We observe that as the guest conducts more searches, the NTC increases. This implies that the model is focusing more on avoiding the negative search milestones as the guest is further into their guest journey.

To understand this further, we analyzed the Pearson correlation coefficients between the historical click-through-rate of a listing and the rejection rate \& cancellation rate of a listing and noticed that the correlations are positive. Recall that click is modeled as a positive search milestone in the Base Module while the rejection and cancellations are modeled as negative search milestones in the Twiddler Module. Thus, a higher NTC for a Twiddler task implies the model is trying to avoid the corresponding negative outcome, and thus indirectly aiming for a lower click-through rate due to the positive correlation between the two. 

In other words, what the model learned aligns with our product intuition. Specifically, we want the guest to explore more in the beginning of the search journey, by showing listings with high click-through rate, even though they could lead to negative outcomes. As the guest narrows down to the final decision, we want to rank higher the listings that will not result in rejection or cancellation.

We hypothesize that the model manages to learn such information exactly due to the different guest behaviors displayed across the search journey. E.g., during the early searches, the guest normally clicks more to better understand the inventory, but narrows down to just a few choices in the latter part of the journey.

\textbf{Summary} - We found that what the Combination Module learns aligns well with our product intuitions in how we might try to balance the guest search journey across guest state and query segment. One might also wonder whether each of the tasks (eg. Base Module output or the three outputs from the Twiddler Module) might already pick up such information through their output score. We plot similar curves for the outputs, \begin{math} y_{t} \end{math} from Equation \ref{equation:combination_equation} in the Combination Module with respect to the same context features and observe that the curves are flat. We hypothesize that the model was unable (or unnecessary) to learn such journey-level information because it has access to both the context features, \begin{math} F_C \end{math}, and the listing features, \begin{math} F_L \end{math}, during the model training. Since the listing features, \begin{math} F_L \end{math} are likely more predictive of the negative search milestones, it dominates the final prediction. By using just the Context Embedding, \begin{math} Emb_C \end{math}, to learn the coefficients in the Combination Module, we essentially guide the module to focus on how these context features could affect the various negative search milestones.

Overall, the ability to interpret what the model learns is extremely helpful from the product perspective, and this helps us to tackle the challenges that we highlighted in Section \ref{section:introduction}.

\subsection{Online Experiments for Stays Ranking}
\label{section:online_experiments}

\begin{table}[]
\begin{tabular}{|c|c|c|ll}
\cline{1-3}
\textbf{Airbnb Product}    & \textbf{Uncancelled Bookers} & \multicolumn{1}{l|}{\textbf{Clickers}} &  &  \\ \cline{1-3}
Stays                      & +0.61\%          & +0.14\%                                &  &  \\ \cline{1-3}
(In-real-life) Experiences & +2.0\%           & +1.1\%                                 &  &  \\ \cline{1-3}
Online Experiences         & +9.0\%           & +1.3\%                                 &  &  \\ \cline{1-3}
\end{tabular}
\caption{Online experiment results across different Airbnb Search Products, all gains have p-values < 0.01}
\label{table:online_results}
\vspace{-6mm}
\end{table}

As shown in Row 1 of Table \ref{table:online_results}, Journey Ranker drove +0.61\% gain in uncancelled bookers compared to the baseline when tested in online A/B experiment. We also saw strong gain throughout the whole conversion funnel, such as an increase of +0.14\% from searchers to clickers, and an increase of +0.48\% from clickers to uncancelled bookers, which aligns with our observation in the offline evaluation.

\subsection{Online Experiments Beyond Stays Ranking}
\label{section:online_experiments_other_search}

In addition to stays ranking, we also successfully applied the Journey Ranker architecture to other search use cases in Airbnb. E.g., row 2 and row 3 in Table \ref{table:online_results} show +2.0\% bookers for experiences ranking and +9.0\% bookers for online experiences ranking. Note that the absolute metric gains for both of these use cases are much higher compared to the Airbnb stays ranking. This is expected since the newer products have a lot less bookings for the model to learn from. Thus, by considering the full guest search journey, we are able to add more incremental value towards better learning guest preference. This also implies that Journey Ranker is most useful for the cold start problem where the absolute number of conversions is smaller, leading to model underfitting, allowing us to more easily ramp up new businesses in the future.

It is worth noting that the implementation for the Experiences use cases is a trimmed down version (i.e. less tasks) of the Airbnb Stays' instantiation described before. This is because the Experiences product lacks reliable telemetry for some intermediate guest actions described. Experiences product also does not have rejection outcomes since all experiences are auto-accepted. Based on offline results for Airbnb stays ranking presented in Table \ref{table:overall_offline_evaluation} and Table \ref{table:base_module_evaluation}, we hypothesize that we can achieve even more gain for the Experiences product once reliable telemetry is added for all of the experience guest actions throughout the search journey.

We also successfully applied the Journey Ranker architecture in Airbnb beyond these search applications. E.g., we collaborated with the Airbnb Email marketing team to leverage the Journey Ranker architecture for the email recommendation use case and achieved +0.7\% nights booked and -3.7\% email unsubscribes.

This is the first modeling effort within Airbnb to achieve significant business gains across multiple Airbnb products with minimal customization thanks to the modular and extensible model architecture. The main considerations needed to leverage this architecture is grouping the guest actions as either multi-step positive milestones in the Base Module or negative milestones in the Twiddler Module and then defining a graded relevance to balance the two.

\section{Conclusion and Future Work}
\label{section:conclusion}
We presented Journey Ranker, a novel multi-task learning framework for Airbnb Stays ranking that balances positive and negative milestones throughout the guest search journey. Journey Ranker significantly outperforms the previous baseline in both offline and online evaluation, and we presented some of the offline study to provide insight into what the model learns. We also saw great results applying Journey Ranker architecture in three other search and non-search use cases in Airbnb. For future work, we believe that other research ideas could easily fit in the Journey Ranker architecture. E.g., we could extend the Base Module with even more micro-tasks that don't follow a sequential multi-step conversion through graph convolution networks \cite{gmcm2020}. We could also easily extend the Twiddler Module with even more negative milestones such as customer support tickets, trip quality etc. Finally, we can modify the Combination Module to directly optimize for business Overall Evaluation Criteria (OEC) and weigh various outcomes using Future Incremental Value (FIV) estimations.

\begin{acks}
No journey (pun intended) is possible without contributions from a lot of people. Specifically, we would like to thank Francisco Selame and Bo Yu for collaborating on the Airbnb Experiences use case; and Teng Wang for the Airbnb Email Marketing use case. We would also like to extend our thanks to Alex Deng, Michelle Du, Anna Matlin, and Tatiana Xifara for their help in the online experiment analysis to enable the deployment of the Journey Ranker into production across multiple Airbnb product use cases. Finally, we would like to thank the entire Airbnb Relevance team for valuable discussions.
\end{acks}

\bibliographystyle{ACM-Reference-Format}
\bibliography{airbnb-multitask}

\end{document}